\documentclass{aastex}

\usepackage{psfig,emulateapj5}

\newcommand{\lsim}{\raisebox{0.3mm}{\em $\, <$} \hspace{-3.3mm}
\raisebox{-1.8mm}{\em $\sim \,$}}
\newcommand{\gsim}{\raisebox{0.3mm}{\em $\, >$} \hspace{-3.3mm}
\raisebox{-1.8mm}{\em $\sim \,$}}

\shorttitle{
Does the Slim-Disk Model Correctly Consider Photon-Trapping Effects?}
\shortauthors{Ohsuga et al.}

\begin{document}


\title{Does the Slim-Disk Model Correctly Consider Photon-Trapping Effects?}


\author{K. Ohsuga and S. Mineshige}
\affil{Yukawa Institute for Theoretical Physics, Kyoto University,
Kyoto 606-8502, Japan}
\author{M. Mori}
\affil{Institute of Natural Science, Senshu University, 
Kawasaki, Kanagawa 214-8580, Japan}
\and
\author{M. Umemura}
\affil{Center for Computational Physics, University of Tsukuba,
Tsukuba, Ibaraki 305-8577, Japan}





\begin{abstract}
We investigate the photon-trapping effects
in the super-critical black hole accretion flows
by solving radiation transfer as well as
the energy equations of radiation and gas.
It is found that the slim-disk model generally
overestimates the luminosity of the disk at around the 
Eddington luminosity ($L_{\rm E}$) 
and is not accurate in describing the effective temperature profile,
since it neglects time delay between energy generation 
at deeper inside the disk
and energy release at the surface.
Especially, the photon-trapping effects are appreciable even below
$L \sim L_{\rm E}$, while they appear above $\sim 3L_{\rm E}$
according to the slim disk.
Through the photon-trapping effects,
the luminosity is reduced 
and the effective temperature profile 
becomes flatter than 
$r^{-3/4}$ as in the standard disk.
In the case that the viscous heating is effective 
only around the equatorial plane,
the luminosity is kept around the Eddington luminosity
even at very large mass accretion rate, $\dot M \gg L_{\rm E}/c^2$.
The effective temperature profile is almost flat,
and the maximum temperature decreases
in accordance with rise in the mass accretion rate.
Thus, the most luminous radius 
shifts to the outer region 
when $\dot{M}/(L_{\rm E}/c^2) \gg 10^2$.
In the case that the energy is dissipated equally at
any heights,
the resultant luminosity is somewhat larger than in the former case,
but the energy-conversion efficiency still decreases 
with increase of the mass accretion rate, as well.
The most luminous radius stays around the inner edge of the disk 
in the latter case.
Hence, the effective temperature profile is 
sensitive to the vertical distribution of energy production rates,
so is the spectral shape.
Future observations of high $L/L_{\rm E}$ objects will be able to
test our model.
\end{abstract}



\section{INTRODUCTION}
X-ray binary sources and active galactic nuclei (AGNs)
emit enormous energy in radiation,
and it is believed that the accretion disks
are the main place of energy release in those objects.
The standard-disk model was proposed by Shakura \& Sunyaev (1973)
as a very efficient 
conversion mechanism of gravitational energy of accreting gas
into radiation energy (Pringle 1981; 
Frank, King, \& Raine 1985; Kato, Fukue, \& Mineshige 1998).
The standard model has been very successful
in describing optically thick flow structure,
as long as the mass accretion rate, $\dot{M}$,
is less than the critical mass accretion rate,
$\dot{M}_{\rm crit} \equiv L_{\rm E}/c^2$,
with $L_{\rm E}$ being the Eddington luminosity given by
$4\pi cGMm_{\rm p}/\sigma_{\rm T}$,
where $c$ is the light velocity,
$M$ is the black hole mass,
$m_{\rm p}$ is the proton mass,
and $\sigma_{\rm T}$ is the Thomson scattering cross-section.
However, sub-critical accretion is not always 
guaranteed,
since the mass accretion rate is determined 
by something other than the central star itself,
i.e.,
companion star in binary system 
(Alme \& Wilson 1976; 
Spruit \& Ritter 1983; 
King et al. 1997; 
Koyama et al. 1999)
or circumnuclear star clusters in AGNs 
(Norman \& Scoville 1988;
Umemura, Fukue, \& Mineshige 
1998; 
Ohsuga et al. 1999).
Therefore, it is no wonder that 
the mass accretion rate can greatly exceed the critical value.

When the mass accretion rate is comparable to or more than
the critical limit,
the disk becomes radiation pressure dominant.
In the case of $\dot{M} \gsim 10 \dot{M}_{\rm crit}$, moreover,
the disk is moderately geometrically thick and
advective energy transport becomes substantial.
Such a disk is called a slim disk and 
has been investigated in detail
(Abramowicz et al. 1988; Szuszkiewicz, Malkan, \& Abramowicz 1996; 
Wang et al. 1999; Watarai \& Fukue 1999; 
Watarai et al. 2000).
The unique feature in high $\dot{M}/\dot{M}_{\rm crit}$ systems
is that 
photons are trapped (Katz 1977; Begelman 1978; 
Begelman \& Meier 1982
; Flammang 1984; Blondin 1986; Colpi 1988; Wang \& Zhou 1999
).
In optically thick accretion flow,
frequent interaction between matter and photons
delays liberation of the radiation energy arising at deeper inside 
the disk.
Thus, the radiation energy is trapped in the flow and advected inward.
Such a photon-trapping plays an important role 
when the radiative diffusion time-scale
is longer than the accretion time-scale.
Since the trapped radiation energy can fall onto the black hole 
with the accreting gas without being radiated away,
the observed luminosity is reduced
in the black-hole accretion flow.
In contrast,
the advected energy should be finally radiated at the stellar surface
in the case of neutron star (Houck \& Chevalier 1992).

While most studies focused on spherical accretion case
with some discussion on Comptonization, we here focus on the disk 
accretion case.
In the slim-disk approach,
the vertically-integrated approximation is adopted; namely
the radiative flux at the disk surface is 
related to the temperature on the equatorial plane, $T_{\rm c}$,
through the relation,
$F^z_{\rm surf} \sim \sigma T_{\rm c}^4/\tau$, 
where 
$\sigma$ is the Stefan-Boltzmann constant
and $\tau$ is the vertical Thomson scattering optical depth 
for a half of the disk.
However, 
this holds
only when 
the radiative diffusion time-scale is shorter than 
the accretion time-scale.
In other words, 
the photon-trapping effects are not exactly taken into consideration
in the slim-disk formulation.
Eggum, Coroniti, \& Katz (1987, 1988) 
initiated the two-dimensional radiation-hydrodynamical simulations 
by assuming the equilibrium between gas and radiation.
Recently, 
the improved simulations, 
in which the energy of gas and radiation are separately treated,
were performed by 
Okuda, Fujita, \& Sakashita (1997), Fujita \& Okuda (1998)
and Kley \& Lin (1999).

In this work, 
by solving radiation transfer,
we investigate the photon-trapping effects,
paying attention to energy transmission inside the disk
in the optically-thick black hole accretion flows.
The validity of the slim-disk model is also 
discussed in terms of the luminosity and the effective temperature profile.
In \S 2, we present basic considerations of 
the photon-trapping effects based on the comparison of 
the accretion time-scale and the radiative diffusion time-scale.
Model and basic equations for numerical simulations are given in \S 3,
and we describe results in \S 4.
Finally, \S 5 and \S6 are devoted to discussion and conclusions.

\section{BASIC CONSIDERATIONS}
The radiation energy generated near the equatorial plane
diffuses toward the disk surface 
at the speed of $\sim c/3\tau$
(Mihalas \& Mihalas 1984),
so that
the time-scale of radiative diffusion is $t_{\rm diff}=H/(c/3\tau)$,
where 
$H$ is the disk half-thickness.
Since the accretion time-scale, $t_{\rm acc}$, is given by $-r/v_r$,
the condition that 
the radiation energy in the disk
is trapped in the flow and falls onto the black hole
is written as $H/(c/3\tau) \gsim -r/v_r$,
where $r$ is the radius and
$v_r$ is the accretion velocity of the flow.
Using the relation, $\tau=\sigma_{\rm T} \Sigma/2m_{\rm p}$, and
the continuity equation of the accretion disk,
$\dot{M} =-2\pi r v_r \Sigma$,
we obtain the photon-trapping radius as
\begin{equation} 
  r_{\rm trap} = \frac{3}{2}\dot{m} r_{\rm g}
  h,
  \label{r_trap}	
\end{equation}
where 
$\Sigma$ is the surface density of the disk,
$\dot{m}$ is the mass accretion rate normalized by 
the critical mass accretion rate, 
$\dot{m} \equiv \dot{M}/\dot{M}_{\rm crit}$,
$r_{\rm g}$ is the Schwarzschild radius, 
and $h$ is the ratio of the half disk-thickness to the radius
$h=H/r$, respectively.
Hence, the photon-trapping effects perform
significant role in the super-critical accretion flow
when $r_{\rm trap}>3r_{\rm g}$, namely at $\dot{m} \gsim 2$,
since $h$ is of the order of unity in radiation pressure dominant region.

The energy dissipated by viscosity at the regions of $r<r_{\rm trap}$
can not be released but is carried onto the black hole,
as long as the radiative diffusion
is the predominant process for energy transport.
Since the radiation arising merely at $r>r_{\rm trap}$ 
can pass through the disk body
to go out from the surface,
the luminosity is roughly estimated by 
\begin{equation}
  L \sim 2 \int^{\infty}_{\max[r_{\rm in},r_{\rm trap}]} 2 \pi r 
  Q_{\rm vis} dr,
\end{equation}
where $r_{\rm in}$ is the inner edge of the disk,
$Q_{\rm vis}$ is the half vertically-integrated viscous heating rate,
and the compressional heating is neglected.
Adopting the viscous hating rate as 
\begin{equation}	
  Q_{\rm vis} \sim \frac{3}{8\pi} \Omega_{\rm K}^2 \dot{M}
  \left[ 1-\left( \frac{r_{\rm in}}{r} \right)^{1/2} \right] 
\label{Qvis}
\end{equation}
(Shakura \& Sunyaev 1973; Lynden-Bell \& Pringle 1974),
with $\Omega_{\rm K}$ being the Keplerian angular speed,
we have in Newtonian approximation
\begin{equation}
  L \sim \frac{r_{\rm g}}{4r_{\rm in}} \dot{M} c^2,
  \label{Lum}
\end{equation}
for $\dot{m}\ll 2r_{\rm in}/(3r_{\rm g}h)$,
and 
\begin{equation}
  L \sim \frac{3}{4} \dot{M}_{\rm crit} c^2
  \left[ 
  \frac{2}{3h}-\frac{2}{3}\left( \frac{2}{3h}\right)^{3/2}
  \left( \frac{r_{\rm in}}{r_{\rm g}\dot{m}} \right)^{1/2}
  \right],
  \label{LumA}
\end{equation}
if $\dot{m}$ is larger than 
$2r_{\rm in}/(3r_{\rm g}h)$.
It is found in the limit of large $\dot{m}$ ($\gg 1$) that 
the luminosity is roughly constant
irrespective of $\dot{m}$ so that
the energy-conversion efficiency,
$\eta \equiv L/\dot{M} c^2
\sim \dot{M}_{\rm crit}/2\dot{M}h=(2h\dot{m})^{-1}$, 
should be remarkably reduced
in the super-critical accretion disk with $\dot{m} \gg 1$.
Also, it is important to note
that the photon-trapping effects 
do not depend on the accretion velocity nor viscosity for a fixed $\dot{m}$,
since the accretion velocity increases if $\alpha$ rises,
but at the same time, the radiative diffusion velocity also increases
in accordance with decrease in the surface density of the disk
(i.e., $\Sigma \propto v_r^{-1}\propto \alpha^{-1}$).

So far,
we have considered the limiting model that the viscous heating occurs
only in the vicinity of the equatorial plane.
Next, we investigate the photon-trapping effects
for the model that 
the gas is heated up in proportion to the density.
Just as in the previous model,
the photon-trapping effects are appreciable within 
the trapping radius,
however, the radiation 
arising near the disk surface is not trapped in the flow.
The condition for the photon trapping to occur at $z$ is given by
$(H-z)/(c/3\tau_z) \gsim -r/v_r$, 
where $z$ is the vertical height and
$\tau_z$ is the vertical optical depth measured from 
the surface of the disk, 
$\tau_z \equiv \int^{\infty}_z \left( \rho \sigma_{\rm T}/m_{\rm p} \right) 
dz$, 
with $\rho$ being the gas density.
Assuming that the gas density is constant in the vertical direction
for simplicity, namely $\tau_z=\tau (H-z)/H$,
and using the continuity equation of the accretion disk as well as
equation (\ref{r_trap}),
we find that the energy dissipated at 
$\tau_z < (r/r_{\rm trap})^{1/2} \tau$
can be radiated away even within the trapping radius.
Therefore, the luminosity
is estimated by
\begin{eqnarray}
  L
  &\sim& 
  2 \int_{r_{\rm in}}^{r_{\rm trap}} 
  2\pi r  Q_{\rm vis} \left( \frac{r}{r_{\rm trap}}\right)^{1/2}
  dr
  +2\int_{r_{\rm trap}}^{\infty} 
  2\pi r  Q_{\rm vis} dr
  \nonumber \\
  &=& \frac{3}{4}\dot{M}_{\rm crit}c^2 
  \left[
  -\frac{2}{3h}+\frac{1}{3}
  \left(\frac{2}{3h} \right)^{3/2}
  \left(\frac{r_{\rm in}}{\dot{m}{r_g}} \right)^{1/2}
  \right.
  \nonumber \\
  & & 
  \left.
  \ \ \ \ \ \ \ \ \ \ \ \ \ \ \ \ \ \ \ \ \ \ \ \
  \ \ \ \ \ \ \ \ \ \ \ \ \ \ \
  +\left( \frac{2\dot{m}r_{\rm g}}{3hr_{\rm in}} \right)^{1/2}
  \right],
  \label{LumB}
\end{eqnarray}
when $\dot{m}> 2r_{\rm in}/(3r_{\rm g}h)$.
In this model, 
the luminosity is not constant but depends on the mass accretion rate
as $\propto \dot{m}^{1/2}$ at $\dot{m}\gg 1$.
The energy-conversion efficiency decreases as $\eta\propto \dot{m}^{-1/2}$.
If $\dot{m}< 2r_{\rm in}/(3r_{\rm g}h)$,
the photon-trapping effects do not appear
and the observed luminosity is given by equation (\ref{Lum}).
%
Here, it is stressed that 
the assessment with using the radiative diffusion velocity is valid 
when the most of radiation energy arises at deeper inside the disk.
If the viscous heating occurs only around the disk surface,
the photon trapping would be ineffective.
The reason of this is that 
the photons arising at near disk surface ($\tau_z \ll 1$) 
can easily escape from the flow
since an average number of the scatterings is $\sim \tau_z$,
in spite of $\tau_z^2$ for photons emitted at $\tau_z\gg 1$
(Sunyaev \& Titarchuk 1985).

In Figure 1, we show the expected luminosity changes as 
functions of $\dot{m}$ assessed from 
equations (\ref{Lum}), (\ref{LumA}), and (\ref{LumB}).
To compare with the numerical results (\S 4.2), 
the former and latter models are represented as
the analytical models A (thick solid curve) and B (dashed curve), 
respectively,
and $h=0.5$ as well as $r_{\rm in}=3r_{\rm g}$ are assumed.
In the sub-critical regime,
the luminosities for both models vary along
the thin solid line.
We also display the luminosity of the slim disk (dotted curve).
These two models are both extreme in the opposite sense,
and actual situation will lie just in between.
We thus expect that the luminosity increases moderately
but rises slowly than $\dot{m}^{1/2}$.
Hence, the slim disk is expected to overestimate the luminosity 
at $L \gsim L_{\rm E}$, namely $\dot{m} \gsim$ a few.
In \S 4.2, we discuss about behavior of the luminosity in more detail.
To assess the radiative flux as a function of radius,
we need to calculate the radius,
at which the energy emerged at outer region is released. 
Note that finite disk size effect enhances photon trapping,
and the disks would be even fainter (see \S 5.2).
\centerline{\psfig{file=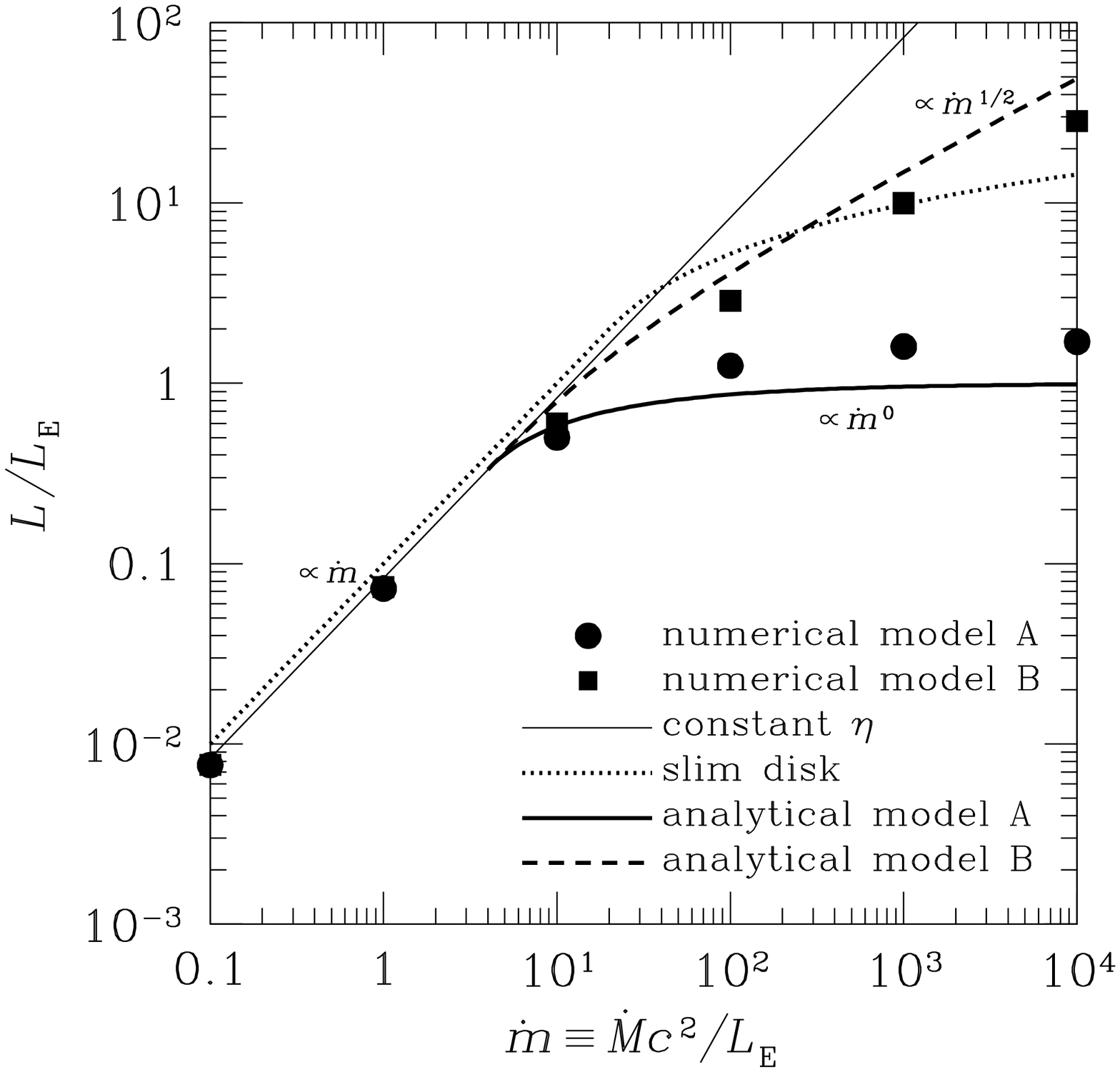,width=3.4in}}
\figcaption[f1.eps] {
The luminosity is plotted against the mass accretion rate,
where the luminosity and mass accretion rate are 
normalized by the Eddington luminosity and the critical mass accretion rate,
respectively.
The thick solid and dashed curves are the luminosities 
analytically estimated in \S 2.
The circles and squares are the numerical results 
for models A and B, respectively.
The luminosity should increase along the thin solid line,
if energy-conversion efficiency is constant.
It is found, however, that 
the energy-conversion efficient is on decrease
in the super-critical accretion flow.
Clearly, the slim disk (dotted curve) considerably overestimates the luminosity
as compared with model A at whole super-critical regime 
and with model B when $\dot{m} \sim 10-10^{3}$.
The numerical results are roughly consistent with 
analytical ones.
\label{fig1}}

\section{MODEL AND BASIC EQUATIONS FOR NUMERICAL SIMULATIONS}
In the previous section, 
we roughly estimate the photon-trapping effects 
by comparing the accretion time-scale and 
the radiative diffusion time-scale.
In this section, we examine the photon-trapping effects
by practically solving radiation transfer,
and the energy equations of gas and radiation.
Since we are more concerned with the photon-trapping effects themselves
rather than flow structure,
we employ a simple model for the accretion flow.
Here, the cylindrical coordinate, $(r, \varphi, z)$, is used.
We consider that the accretion disk is axisymmetric 
and steady in the Eulerian description; 
$\partial/\partial \varphi = \partial/\partial t = 0$.
The radial component of the velocity 
is expressed in terms of free-fall velocity as
\begin{equation}
   v_r = - \xi \left( \frac{GM}{r} \right)^{1/2},
  \label{vr}	
\end{equation}
where $\xi$ is a constant parameter,
and its vertical component is prescribed as
\begin{equation}
   v_z = \frac{z}{r} v_r,
  \label{vz}	
\end{equation}
i.e., we assume convergence flow.
Note that $v_r$ is related to the viscosity parameter $\alpha$ 
through $v_r \sim \alpha \left( H/r \right) c_{\rm s}$,
where 
$c_{\rm s}$ is the sound velocity.
We focus on the accreting ring element 
whose geometrical width and thickness are $\Delta r$ and $2H$,
respectively, where $\Delta r \ll r$ (see Figure 2).
We suppose the structure of the accretion disk to be 
locally plane parallel,
that is to say, each ring is composed of $N$ layers in the $z$-direction,
and the thickness of each layer is $\Delta z = 2H/N$.
We solve the time-dependent energy fields of gas and radiation
in the ring element during the course of accretion motion
until the element reaches the inner edge of the disk.
We adopt uniform density profile,
\begin{equation}
  \rho(r, z) = \frac{\Sigma (r) }{2H (r)}=\mbox{const. in $z$},
  \label{density}	
\end{equation}
and consider both of the viscous heating and compressional heating 
due to converging inflow.
Since the flow is steady ($\partial/\partial t = 0$),
time coordinate can be transformed to the spatial coordinates;
i.e., $D/Dt=v_r\partial/\partial r+v_z\partial/\partial z$
with $v_r$ and $v_z$ being given by equations (\ref{vr}) and (\ref{vz}).
We thus express the radiative flux at the disk surface 
as a function of radius 
and then calculate the resultant luminosity of the disk.
\centerline{\psfig{file=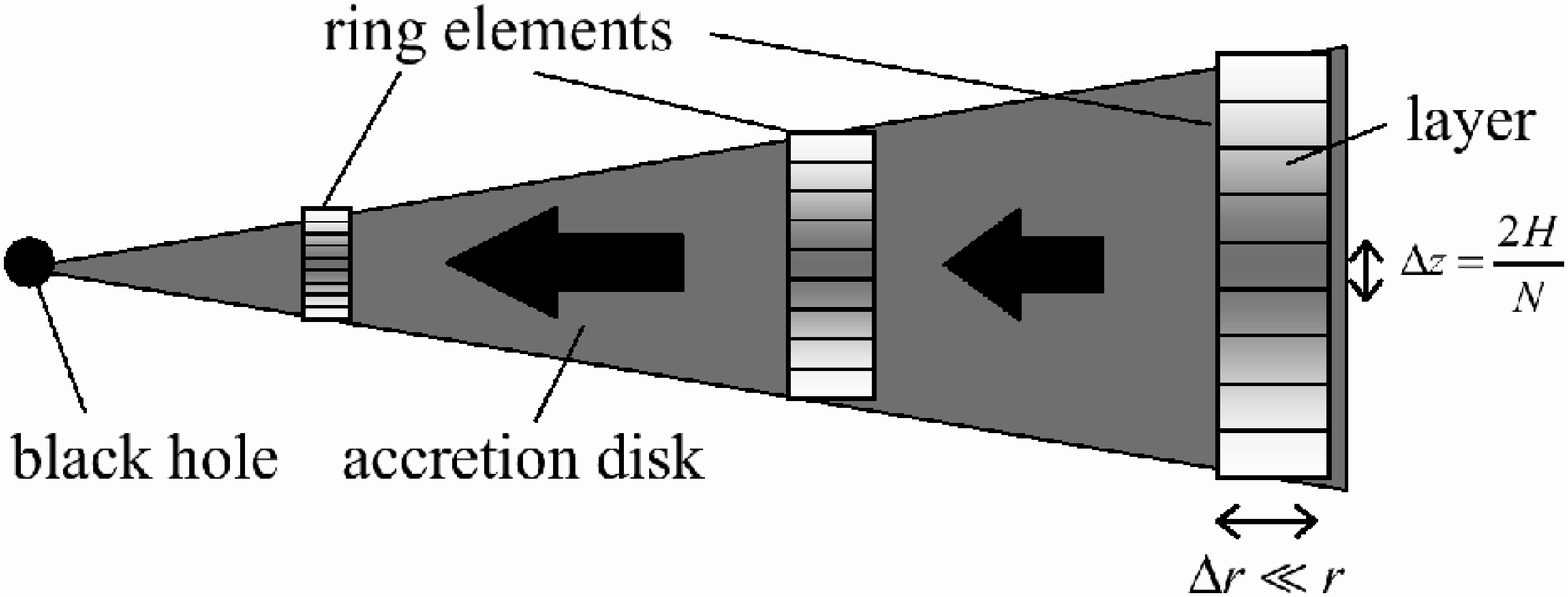,width=3.4in}}
\figcaption[f2.eps] {
Schematic view of the black-hole accretion system
explaining our calculation method.
We calculate the evolution of the energy fields in a moving ring element
that consists of $N$ layers,
until the element reaches the inner edge of the disk,
assuming layers in each ring to be locally plane-parallel.
Since the steady flow in the Eulerian description is considered,
we set $\partial/\partial t = 0$,
and thus replace $D/Dt = v_r \partial/\partial r+v_z\partial/\partial z$.
\label{fig2}}
%
%
%
%
%


\medskip
In plane parallel approximation,
the radial and azimuthal components of the radiative flux 
are null.
Also, non-diagonal components of the radiation stress tensor
are null.
Hence, using equations (\ref{vr}) and (\ref{vz}), 
we write the energy equations of radiation and gas as 
\begin{eqnarray}
   \rho \left( v_r \frac{\partial}{\partial r} \right.
   &+& \left. \frac{z}{r} v_r \frac{\partial}{\partial z} \right)
   \left(\frac{E}{\rho}\right)
   \nonumber \\
   &=& -\frac{\partial F^z}{\partial z}
   +\frac{v_r}{r} 
   \left( \frac{1}{2} P^{rr}
   -P^{\varphi\varphi}
   -P^{zz}
   \right)
   \nonumber\\
   & & +4\pi \kappa B - c\kappa E,
\label{radene1}
\end{eqnarray}
and 
\begin{equation}
   \rho \left( v_r \frac{\partial}{\partial r}
   +\frac{z}{r} v_r \frac{\partial}{\partial z} \right)
   \left(\frac{e}{\rho}\right)
   = -\frac{3}{2}\frac{v_r}{r}p_{\rm gas}  -4\pi \kappa B + c\kappa E
   + q_{\rm vis},
\label{gasene}
\end{equation}
respectively,
where 
$E$ is the radiation energy density,
$F^z$ is the radiative flux in the $z$-direction,
$P^{ii}$ is the diagonal components of the radiation stress tensor,
$\kappa$ is the absorption coefficient,
$B$ is the blackbody intensity written as
$B=\sigma T_{\rm gas}^4/\pi$ with $T_{\rm gas}$ 
being the gas temperature,
$e$ is the internal energy density,
$p_{\rm gas}$ is the gas pressure,
and $q_{\rm vis}$ is the viscous heating rate per unit volume
(Mihalas \& Klein 1982; Mihalas \& Mihalas 1984; 
Fukue, Kato, \& Matsumoto 1985;
Stone, Mihalas, \& Norman 1992).
%
To close the set of equations,
we apply the flux-limited diffusion approximation
for the radiative flux and stress tensor,
\begin{equation}
   F^z = - \frac{c\lambda}{\chi}\frac{\partial E}{\partial z},
\end{equation}
\begin{equation}
   P^{rr} = P^{\varphi\varphi} = \frac{1}{2}(1-f)E,
\end{equation}
and 
\begin{equation}
   P^{zz} = fE,
\end{equation}
with $\chi$ being the extinction coefficient,
where $\lambda$, $R$, and $f$ are defined as 
\begin{equation}
   \lambda = \frac{2+R}{6+3R+R^2},
\end{equation}
\begin{equation}
   R = \frac{1}{\chi E} \left| \frac{\partial E}{\partial z} \right |,
\end{equation}
and 
\begin{equation}
   f=\lambda + \lambda^2 R^2
\end{equation}
(Turner \& Stone 2001).
This approximation holds
both in the optically thick and thin regimes.
In the optically thick limit, we find
$\lambda \rightarrow 1/3$ and $f \rightarrow 1/3$
because of $R \rightarrow 0$.
In the optically thin limit of $R\rightarrow \infty$,
on the other hand, 
we have $| F^z | = cE$, $P^{rr}=P^{\varphi\varphi}=0$,
and $P^{zz}=E$.
These give correct relations in the optically thick diffusion limit
and optically thin streaming limit, respectively.
Thus, we can rewrite the radiation energy equation (\ref{radene1}) as
\begin{eqnarray}
   \rho & & \! \left( v_r \frac{\partial}{\partial r} 
   +\frac{z}{r} v_r \frac{\partial}{\partial z} \right)
   \left(\frac{E}{\rho}\right)
  \nonumber \\
   & &\!\!\!=c\frac{\partial}{\partial z} \left(
   \frac{\lambda}{\chi}\frac{\partial E}{\partial z} \right)
   -\frac{3f+1}{4}\frac{v_r}{r}E
   +4\pi \kappa B - c\kappa E.
\label{radene}
\end{eqnarray}

These nonlinear equations (\ref{gasene}) and (\ref{radene}) are integrated 
iteratively by the Newton-Raphson method 
with the Gauss-Jordan  elimination for a matrix inversion,
by coupling with the continuity equation,
\begin{equation}
  \dot{M}= -2\pi r v_r \Sigma,
\end{equation}
the density profile given by equation (\ref{density}),
and the equation of state,
\begin{equation}
   p_{\rm gas} = \frac{2}{3}e.
\end{equation}
As to the vertical distribution of viscous heating rates,
we consider two extreme models.
In model A, we assume that 
the viscous heating is effective only in the vicinity of the
equatorial plane at $|z|<10^{-2}H$, that is to say,
\begin{equation}
  q_{\rm vis}(z) = \left\{ 
  \begin{array}{ll}
    10^2 Q_{\rm vis}/H & |z|\le 10^{-2}H \\
    0 & |z|>10^{-2}H
  \end{array}
  \right. ,
\end{equation}
where 
$Q_{\rm vis}$ is given by equation (\ref{Qvis}).
In model B, 
we assume 
that the gas is heated up uniformly, independently of $z$,
that is,
\begin{equation}
  q_{\rm vis}(z) = \frac{Q_{\rm vis}}{H}={\rm const}.
\end{equation}
Throughout the present study,
the black hole mass is fixed to be $10 M_\odot$ 
and the inner edge of the disk is taken to be at 
$r_{\rm in}=3r_{\rm g}$.
Moreover, we assume $H$ to be $0.5 r$ (i.e., $h=0.5$),
and $N=100$ is employed.
This assumption of $H=0.5r$ would be reasonable
in the super-critical accretion regime,
since $H/r$ of the radiation pressure dominant disk
becomes of the order of unity.
In the sub-critical regime, on the other hand,
$H$ would be much smaller than $r$,
but then the most of emergent energy can be released immediately
without being trapped.
Thus, our numerical simulations are valid
in both super-critical and sub-critical accretion regimes
for evaluating the radiative flux at the disk surface.

\section{RESULTS}
\subsection{Reduction of the radiative flux}
The radiative flux is reduced due to the photon trapping
in the vicinity of the inner edge of the disk when $\dot{m} \gg 1$.
Figure 3 shows the ratios of the radiative flux at the disk surface,
$F^z_{\rm surf}$,
to the vertically-integrated viscous heating rate
for several $\dot{m}$ as functions of radius.
The solid and dashed curves correspond to model A and model B,
respectively.
Here, we consider the situation that 
the Thomson scattering is predominant over the absorption,
$\kappa=10^{-2}\rho\sigma_{\rm T}/m_p$.
We set $\xi$ in equation (\ref{vr}) for different $\dot{m}$
to be $(\dot{m}, \xi) = (0.1, 10^{-4}), (1, 10^{-3}), (10, 10^{-3}),
(10^2, 10^{-2}), (10^3, 10^{-2})$, and $(10^4, 0.1)$
in order for the disk to be optically thick for Thomson scattering.
Then, the viscosity parameter $\alpha$ is of the order of $10^{-4}$ 
in $\xi = 10^{-4}$ and $0.1$ in $\xi = 0.1$,
since $\alpha \sim \xi/h^2$ from 
equations (\ref{vr}) and (\ref{vz}).
As shown in Figure 3,
the ratio of $F^z_{\rm surf}/Q_{\rm vis}$ 
is around unity 
when the accretion rate is 
sub-critical, $\dot{m} \lsim 1$.
In the super-critical accretion disk, $\dot{m} \gg 1$,
on the other hand, the ratio is much smaller than unity 
in the vicinity of the inner edge of the disk,
but is still close to unity at the outer region.
The ratio rises at $r \sim 3r_g$ again,
but this rise is caused by the decline of $Q_{\rm vis}$ 
near the inner boundary.
The absolute value of the radiative flux is still suppressed there.
\centerline{\psfig{file=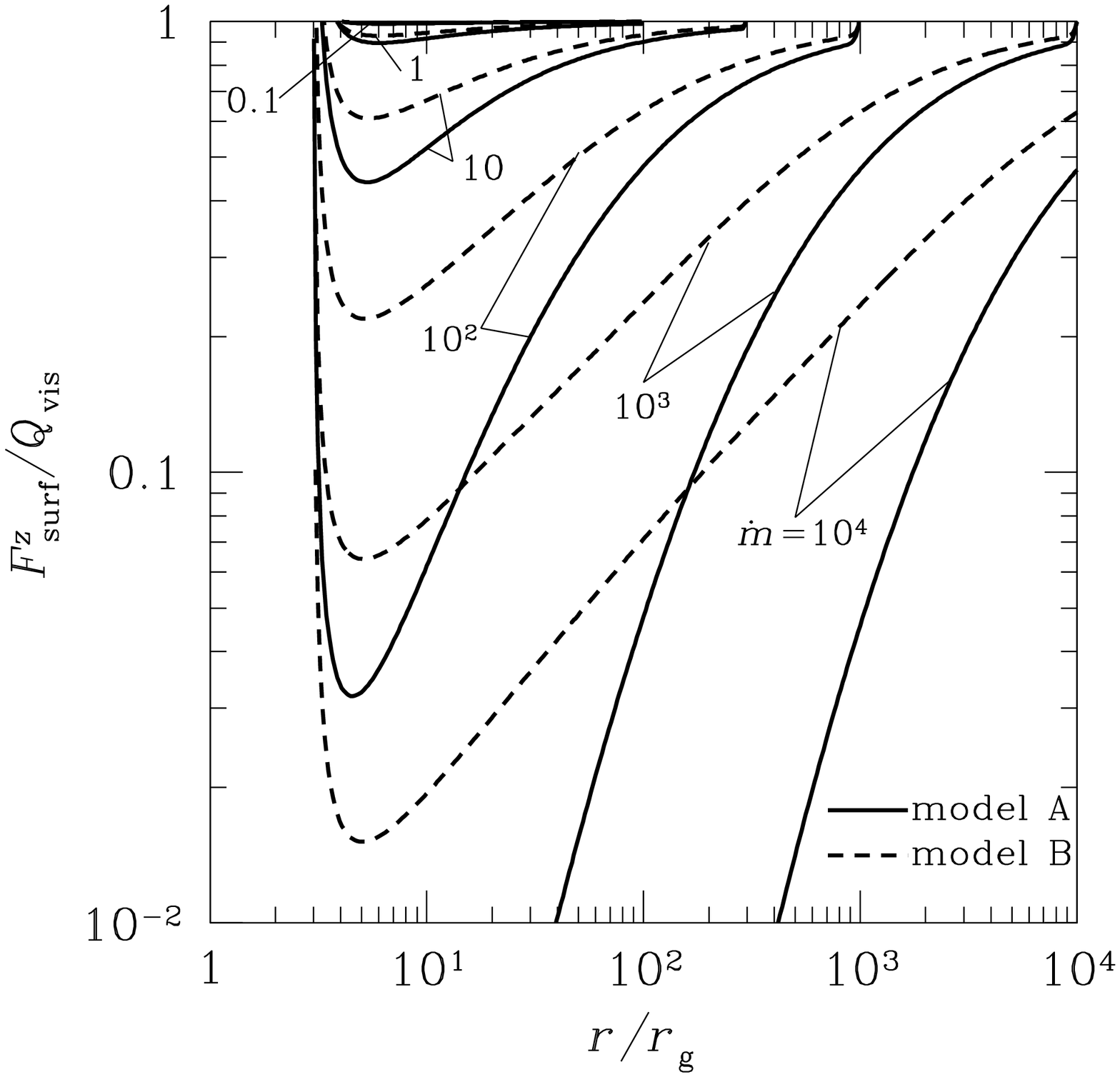,width=3.4in}}
\figcaption[f3.eps] {
The ratios of radiative flux at the disk surface 
to the vertically integrated 
viscous heating rate as functions of radius
for $\dot{m} = 0.1, 10, 10^2, 10^3$, and $10^4$.
Here, the radius is normalized by the Schwarzschild radius.
The solid and dashed curves indicate models A and B, respectively.
By the photon-trapping effects,
the ratios are
much smaller than unity 
at the regions of $r<r_{\rm trap}$ in the case of $\dot{m} \gg 1$,
where $r_{\rm trap}$ is the photon-trapping radius
given by equation (\ref{r_trap}).	
It is also found that the photon-trapping effects are more conspicuous 
in model A than in model B.
If $\dot{m} \leq 1$,
on the other hand, 
we found $F^z_{\rm surf}/Q_{\rm vis} \sim 1$, as in the standard disk.
}

\medskip
This reduction of the radiative flux is due to the photon trapping,
and it can be understood by the time delay owing to
energy transport in the optically thick medium.
The emergent radiation energy is diffused at the speed of $\sim c/3\tau$,
and arrives at the disk surface after $t_{\rm diff}$.
If this time delay is comparable to or longer than the accretion time-scale, 
$t_{\rm diff} \gsim t_{\rm acc}$,
the generated radiation energy is advected inward,
so that the ratio of $F^z_{\rm surf}/Q_{\rm vis}$ should deviate from unity.
To conclude,
the energy dissipated by viscosity
can be released from the surface immediately 
at the same radius, as in the standard disk,
only in the outer region for $\dot{m} \gsim 1$
but in the whole region for $\dot{m} < 1$,
where the radiative diffusion time-scale is much shorter than
the accretion time-scale.

As shown in Figure 3,
the photon-trapping effects are appreciable within $\dot{m} r_{\rm g}$ 
in both models A and B.
This is good agreement with the analytical prediction of $r_{\rm trap}$,
[see equation (\ref{r_trap})].
However, the photon-trapping effects 
are more conspicuous in model A than in model B,
since in model B the energy dissipated near the disk surface
can still escape from the accretion flow 
even within the trapping radius.
In model A, in contrast, 
most of energy is dissipated in the vicinity of the equatorial plane
and is trapped in the flow.
Realistic situation will be just in between.

Here, it should be stressed again that 
the photon-trapping effects themselves depend on $\dot{m}$ 
and are independent of $v_r$ when $\dot{m}$ is fixed,
as was mentioned in \S 2,
unless the disk is optically thin.
To check if this is the case,
we examine the cases with different $\xi$ values; $\xi=10^{-3}$ and $0.1$,
besides $\xi=10^{-2}$, for $\dot{m}=10^2$,
finding no changes.
[However, if the absorption coefficient is very small,
the luminosity would decrease in the case of rapid accretion,
since thermal energy of the gas can not be converted 
into radiation energy (see \S 4.6).]

\subsection{Luminosity}
The luminosity of the disk 
is significantly reduced by the photon-trapping effects.
In Figure 1, the calculated luminosity is also plotted against
$\dot{m}$ with symbols.
Here, the circles and squares indicate the luminosities for models A
and B, respectively. 
The thin solid line represents the luminosity expected 
on the assumption of the constant energy-conversion efficiency ($\eta$).
In model A, 
the luminosity is kept around the Eddington luminosity 
irrespective of the mass accretion rate
in the super-critical accretion regime,
although the luminosity is in proportion to the mass accretion rate
in the sub-critical accretion disk.
Thus, the energy-conversion efficiency is 
remarkably reduced in the super-critical accretion 
due to significant photon-trapping effects, 
$\eta \sim \dot{m}^{-1}$.
The calculated luminosity in model B 
in the super-critical regime
is larger than in model A 
owing to relatively ineffective photon trapping.
Even so, the energy-conversion efficiency is also categorically
on decrease in the super-critical accretion regime,
$\eta \propto \dot{m}^{-1/2}$.
In both models, the results are in good agreement with the
analytical estimates (\S2).

The dotted curve is the luminosity calculated based on
the slim-disk model (Watarai, Mizuno, \& Mineshige 2001).
The energy is assumed to be mainly dissipated on 
the equatorial plane in the slim disks,
however, the luminosity of the slim disk
is not kept around the Eddington luminosity but
increases with the mass accretion rate.
Hence, the slim disks considerably overestimate the luminosity
compared with our similar model (model A).
The luminosity is also 
larger in the slim disk than in model B
in the region of $\dot{m}=10-10^3$.
The photon-trapping effects are already appreciable at $L\lsim L_{\rm E}$,
while they are only substantial at $L\gsim L_{\rm E}$ in the 
slim-disk treatment. 
The slim-disk model thus underestimates the photon trapping
(see \S 1).

Numerical results are roughly consistent with the analytical
assessments plotted based on equations (\ref{LumA}) and (\ref{LumB}),
but deviations originate for the following reason.
In numerical simulations,
the energy transport at each height is actually solved by
taking radiation transfer into consideration.
On the other hand, the analytical assessments are merely obtained by
the comparison of the accretion time-scale
and the radiative diffusion time-scale.

\subsection{Effective temperature}
Since photon trapping tends to lower $F^z_{\rm surf}$ at smaller radii,
the disk with super-critical accretion rate 
has a flatter effective temperature profile 
in the vicinity of the inner edge.
We calculate the effective temperature,
$T_{\rm eff} \equiv \left[ F^z_{\rm surf}/\sigma \right]^{1/4}$,
and plot in Figure 4
the effective temperature distributions 
for model A (solid curves) and model B (dashed curves), respectively.
Although we adopt $M=10M_\odot$,
the results can be applied to cases with other $M$,
if we vary the absolute value of $T_{\rm eff}$ 
in proportion to $M^{-1/4}$.
The effective temperature
for low $\dot{m}$, $\dot{m}=0.1$ and $1$, 
is proportion to $r^{-3/4}$ as in the standard disk 
(Shakura \& Sunyaev 1973; Pringle 1981),
since the photon-trapping effects are not appreciable
when the mass accretion rate is sub-critical (see Figure 3). 
In the case of the super-critical accretion disk,
the slope of the effective temperature profile 
becomes flatter within the trapping radius,
though we still have $T_{\rm eff} \propto r^{-3/4}$ 
at $r>r_{\rm trap}$. 
For $\dot{m} > 10^3$ in model A,
especially, the profile within the trapping radius is almost flat 
and the maximum temperature begins to decrease 
with farther rise in $\dot{m}$.
This makes a marked difference from the prediction by the slim disk,
which shows $T_{\rm eff} \propto r^{-1/2}$
in the high $\dot{m}$ limit
(Watarai \& Fukue 1999; Watarai et al. 2000; discussed later)
\centerline{\psfig{file=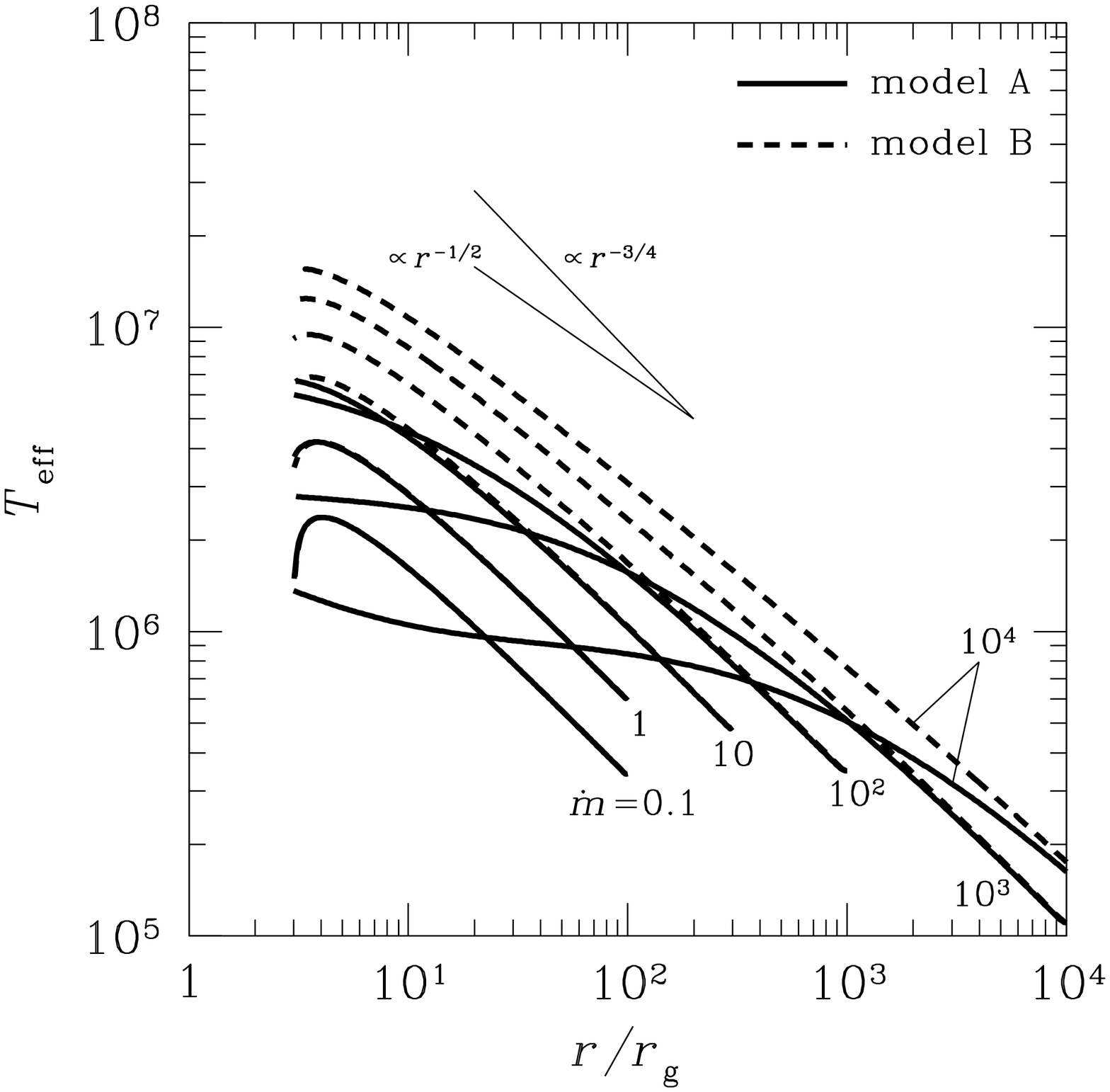,width=3.4in}}
\figcaption[f4.eps] {
The effective temperature profile
for model A (solid curves) and
for model B (dashed curves) for various values of $\dot{m}$.
In the case of $\dot{m} \leq 1$,
the effective temperature is proportion to $r^{-3/4}$,
as in the standard disk.
However, the profile drastically changes 
in the super-critical accretion flow ($\dot{m} \gg 1$),
and becomes significantly flatter at the region of $r<r_{\rm trap}$.
In model A, the maximum temperature decreases with
increase in the mass accretion rate
and the profile is nearly flat when $\dot{m} > 10^3$.
}

\subsection{Spectral index}
As we have seen above, the effective temperature profile
is very sensitive to the functional form of $q_{\rm vis}(z)$.
The vertical distribution of viscous heating rates would be observationally 
distinguished 
when the luminosity is comparable to or more than the Eddington luminosity.
The spectral energy distribution (SED) is strongly affected by
the effective temperature profile, which in turn depends on $q_{\rm vis}(z)$.
We have calculated expected SED based on a superposition of blackbody spectra, $B_\nu[T_{\rm eff} (r)]$,
emitted from the disk surface, and plot in Figure 5
the spectral index, $\zeta = d\log L_\nu /d\log \nu$,
in the range between 0.1 keV and 1.0 keV,
as a function of the luminosity normalized by the
Eddington luminosity.
Here, the circles and squares indicate the spectral indices for models A
and B, respectively. 
It is known that the spectrum of the standard type disk but with
$T_{\rm eff} \propto r^{-p}$ is composed of three components:
Rayleigh-Jeans part ($L_\nu \propto \nu^2$) at low $\nu$,
Wien part [$\propto \nu^3 \exp(-h\nu/kT_{\rm eff})$] 
at high $\nu$, and 
a power-law part ($\propto \nu^{3-2/p}$) in the intermediate frequencies
(Kato, Fukue, \& Mineshige 1998),
where $\nu$ is the photon frequency,
$h$ is the Planck constant,
and $k$ is the Boltzmann constant.
Thus, difference in $p$ should manifest in this middle part.
\centerline{\psfig{file=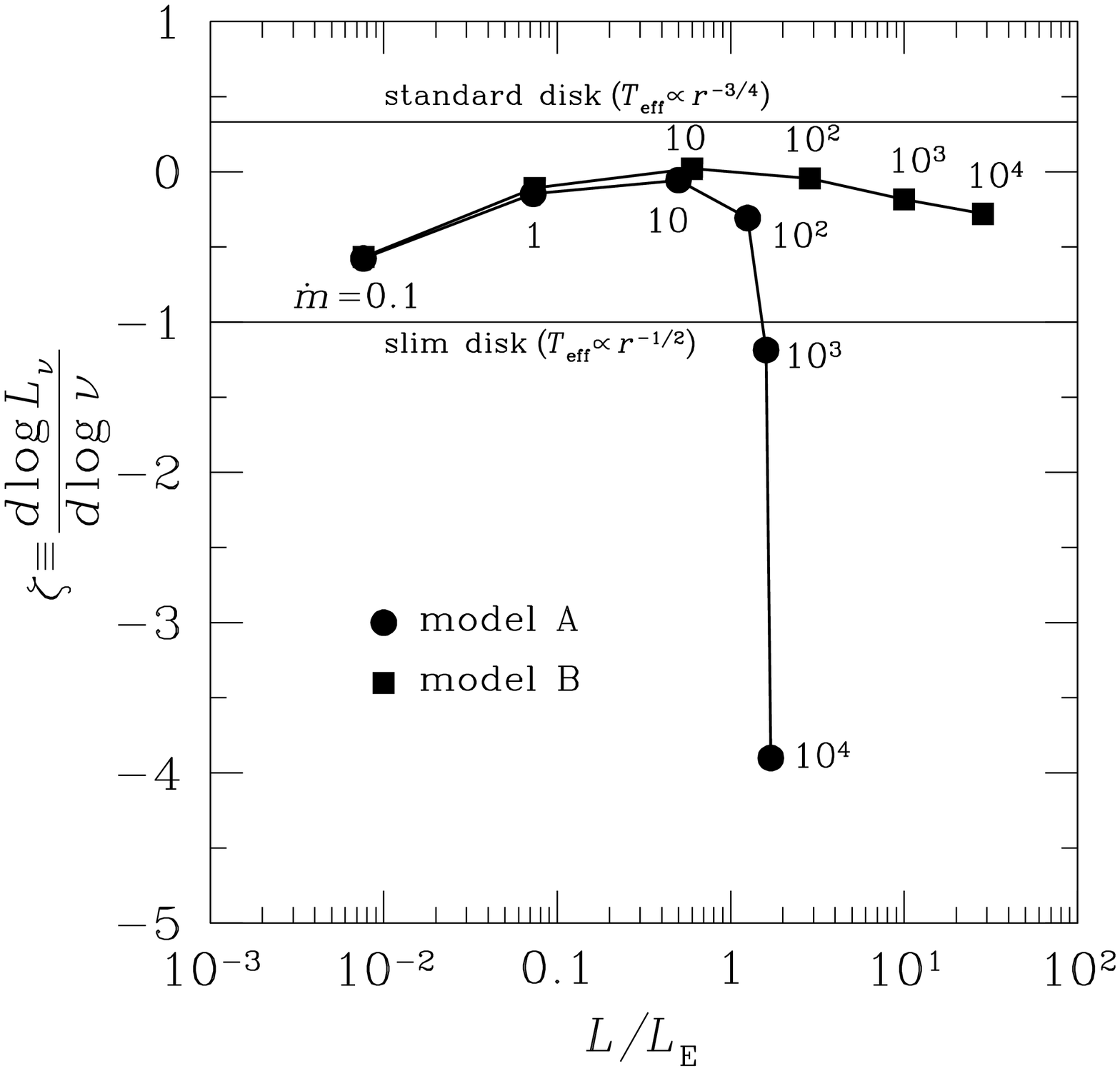,width=3.4in}}
\figcaption[f5.eps] {
The spectral indices 
as functions of normalized luminosity, $L/L_{\rm E}$.
Here, we define the spectral index as 
$\zeta = d\log L_\nu / d\log \nu$ at 0.1--1.0 keV.
The circles and squares correspond to models A and B, 
respectively.
The spectral index is sensitive to
the distributions of the viscous heating rates in
the vertical direction, 
thus it can be determined observationally.}
\medskip
The effective temperature for $\dot{m}=0.1$ and $1$
is proportional to $r^{-3/4}$,
however, the index deviates from $\zeta=1/3$ for $p=-3/4$ 
expected from the standard-disk model,
since we see a Wien cut-off on the high-frequency side.
The effective temperature profile in model A becomes flatter 
and the maximum temperature begins to decrease when $\dot{m} > 10^2$,
so that the peak of the SED shifts to the low-frequency side.
Thus, the index drastically decreases in accordance with 
increasing $\dot{m}$, 
although the luminosity stays around the Eddington luminosity.
In model B, on the other hand, the maximum temperature continues to increase
with rise in the mass accretion rate.
Therefore, the Wien cut-off is out of the range
in the super-critical accretion.
As shown in Figure 4, the effective temperature profile in model B
is flatter than in the standard disk, $\propto r^{-3/4}$,
but steeper than in the slim disk of high $\dot{m}$ limit, $\propto r^{-1/2}$.
Therefore, the index is settled down between $1/3$ (standard disk)
and $-1$ (slim disk), and 
the luminosity increases in accordance with rise in the mass accretion rate.
We can thus obtain information of $q_{\rm vis}(z)$ by 
the X-ray observations of luminous X-ray objects.

\subsection{Most luminous radius}
The radius, at which differential luminosity
[$(dL/dr) \Delta r$] is maximum,
is at $(75/16)r_{\rm g}$ in the standard disk.
This 'most luminous radius' does not necessarily coincide
with the inner edge radius, if the photon trapping is effective,
since it is more prominent at small radii.
We show the most luminous radii, $r_{\rm max}$, 
normalized by $r_{\rm g}$ in Figure 6.
The circles and squares indicate $r_{\rm max}/r_{\rm g}$ for 
model A and model B, respectively.
This figure clearly demonstrates for model A that the larger $\dot{m}$ is,
the larger becomes the most luminous radius,
as long as $\dot{m} \gg 10^2$.
In model B,
the most luminous radius stays at around the inner edge of the disk,
but the photon trapping significantly reduces the overall radiative flux 
within the trapping radius.
\centerline{\psfig{file=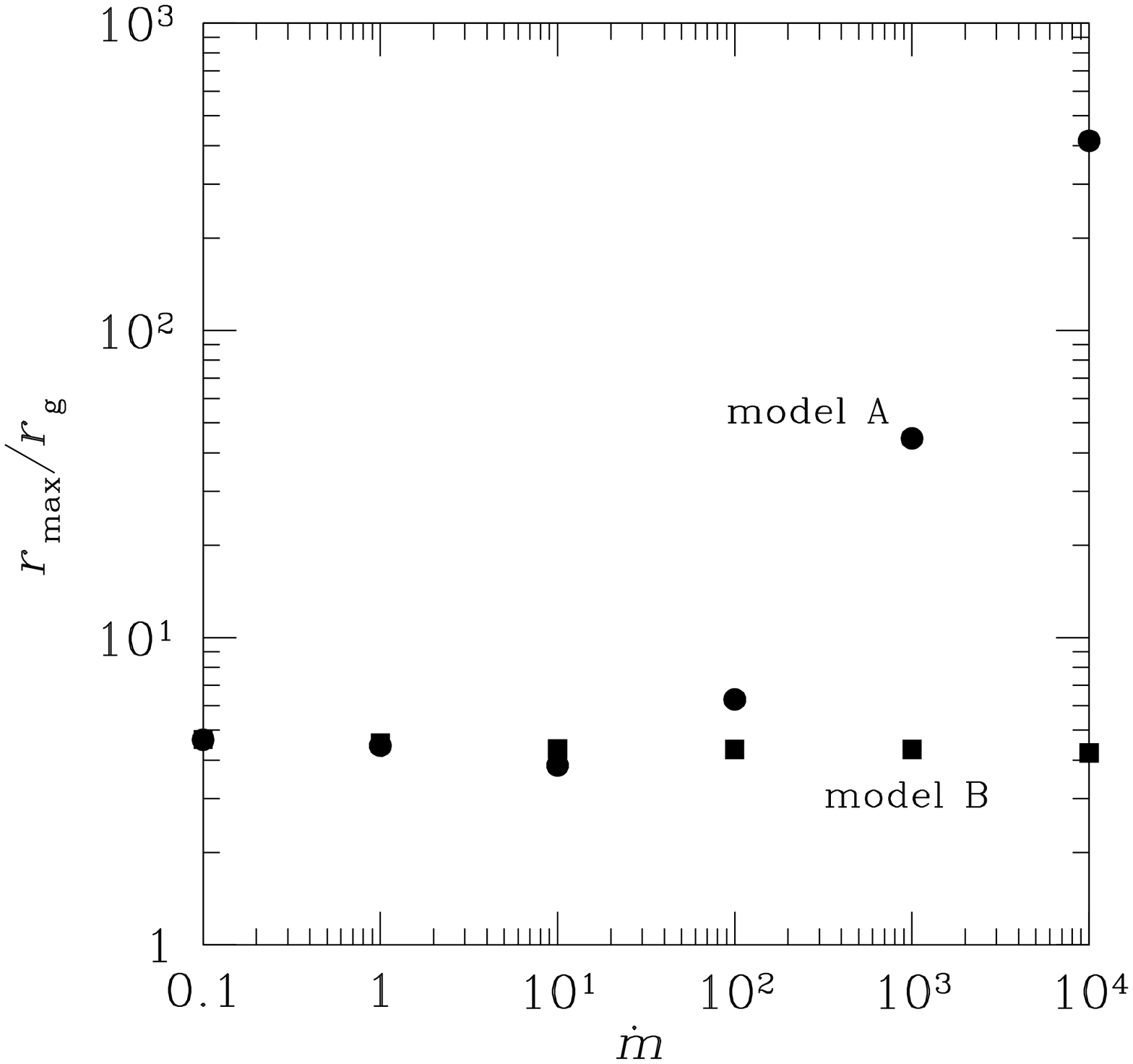,width=3.4in}}
\figcaption[f6.eps] {
The most luminous radii,
at which $(dL/dr) \Delta r$ is maximum, 
normalized by $r_{\rm g}$ are shown 
as functions of normalized mass accretion rate, $\dot{m}$.
The circles and squares correspond to models A and B, 
respectively.
Since the photon-trapping effects are most appreciable
in the vicinity of the inner edge,
the most luminous radius shifts 
from the inner edge to the outer region at high $\dot{m}$
in model A.
In model B, in contrast, the innermost ring is still most luminous,
although the radiative flux at the surface is reduced
within $r_{\rm trap}$ (see Figure 3).
}

\subsection{Rapid accretion}
So far, 
since we adopt relatively large absorption coefficient,
the gas is approximately in equilibrium with radiation, 
$T_{\rm gas} \sim T_{\rm rad}$,
where $T_{\rm rad}$ is the radiation temperature.
Thus, the energy dissipated by viscosity
can be promptly converted into radiation energy, 
and that energy is transported toward the disk surface by radiation transfer.
However, if the absorption coefficient
is much smaller, 
as in the case that we adopt free-free absorption,
it is possible that the energy dissipated by viscosity
is kept as thermal energy and is not converted into radiation energy.
Therefore, the resultant luminosity must further decrease
because of this effect
in addition to the photon-trapping effects,
if the gas accretes onto the black hole without 
achieving radiative equilibrium.
This occurs, when $\alpha$ is relatively large $\alpha \gsim 0.03$
(Beloborodov 1998).

To demonstrate this phenomenon,
we employ the Rosseland mean of the free-free absorption coefficient,
$\kappa_{\rm ff} = 1.7\times 10^{-25} T_{\rm gas}^{-7/2}
(\rho/m_{\rm p})^2$
(Rybicki \& Lightman 1979),
and calculate and plot 
the luminosity against the radial velocity for model A (circles)
and model B (squares) in Figure 7, where the mass accretion rate is 
fixed to be $\dot{m}=10^2$.
It is found that the accretion disk gets much fainter at the regime 
of $\xi=v_r/v_{\rm ff} > 10^{-2}$
[corresponding to $\alpha \gsim 0.04$ if $H/r\sim 0.5$,
since $v_r \sim \alpha (H/r)^2 v_\phi$],
where $v_{\rm ff}$ is the free-fall velocity.
The accretion disks with huge accretion rate and
large radial velocity tend to have extremely small 
energy-conversion efficiency 
and would be identified as faint objects.
Note, however, that 
this criterion of the radial velocity might vary,
depending on the vertical density profile,
since the free-free absorption coefficient strongly depends on the
gas density, as well as the gas temperature.
If the density and temperature of the gas is high around the equatorial plane,
the thermal energy 
would be converted into the radiation energy there,
and the disk might be somewhat brighter.
Anyhow, the disk tends to be fainter in more rapid accretion
in cooperation with the photon-trapping effects,
although the detail examination requires 
multi-dimensional radiation-hydrodynamical simulation.
(This will be performed in future.)
\centerline{\psfig{file=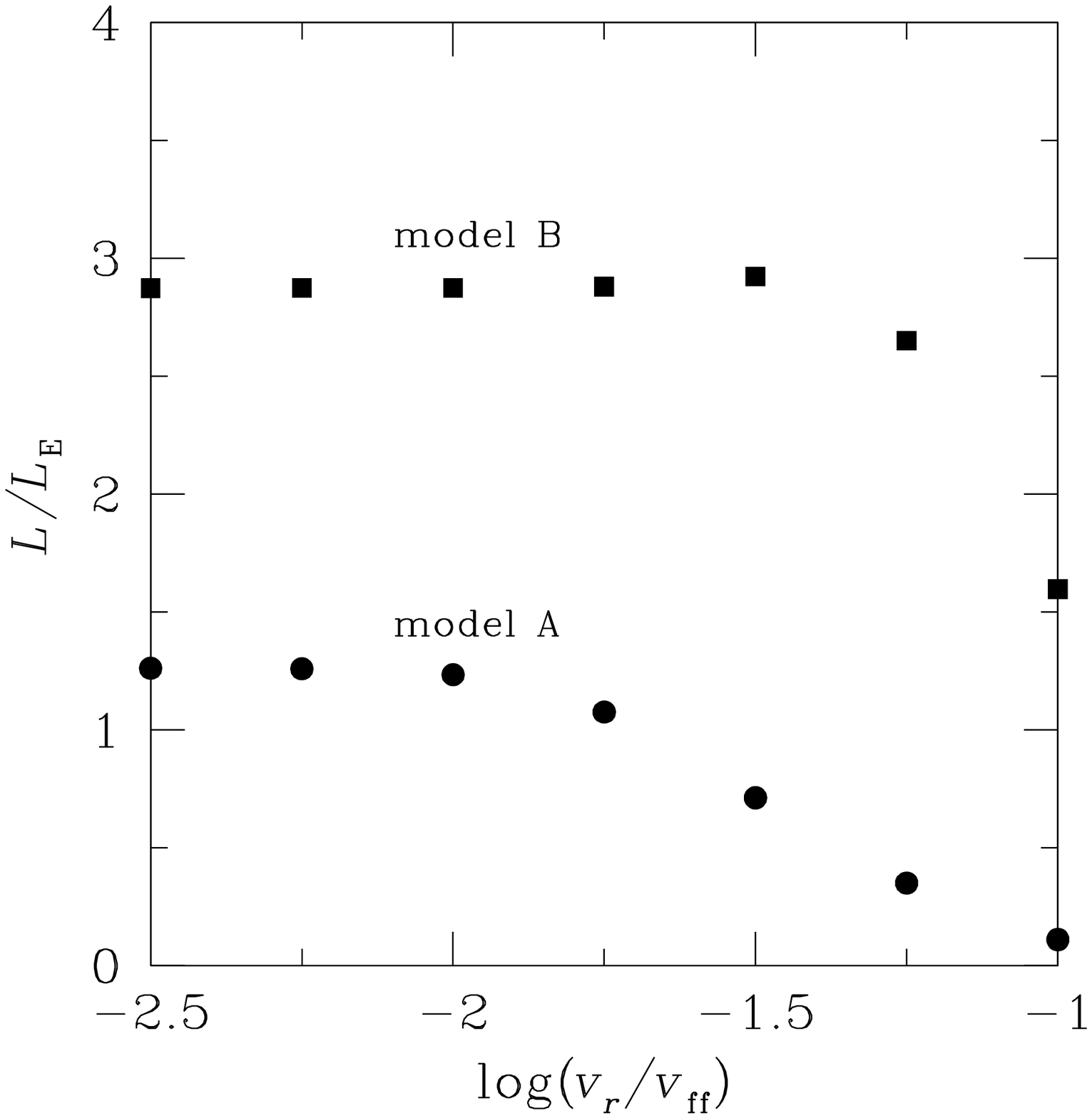,width=3.4in}}
\figcaption[f7.eps] {
The luminosity of the flow as a function of the accretion velocity,
where the circles and squares indicate model A and model B, 
respectively.
Here, we adopt the Rosseland mean of the free-free absorption coefficient.
The accretion disk becomes dimmer
in the regime of $v_r > 10^{-2}v_{\rm ff}$ $(\xi > 10^{-2})$,
since the energy dissipated by viscosity is not 
sufficiently converted into radiation energy but kept as 
thermal energy of the gas.
The luminosity of the accretion disk is suppressed 
by this effect as well as the photon-trapping effects
in the case of the disk with rapid accretion
(i.e., large $\alpha\geq 0.04$).}

\section{DISCUSSION}
\subsection{Comparison with the slim-disk model}
As a model for describing super-critical disk accretion flow,
slim disks are often utilized.
However, we wish to note that the slim disk does not accurately
treat the photon-trapping effects in the disk accretion case.
The slim disk can describe the dynamical structure of the flow 
reasonably well,
since the energy balance is given by 
$Q_{\rm vis} \sim Q_{\rm adv}$ and does not depend on the 
radiative loss. 
The slim disk is not able to correctly estimate 
the radiative flux at the disk surface,
in the sense that it overestimates the luminosity in all ranges
in comparison with model A, 
and in the range of $\dot{m}$ between $10$ and $10^3$ in model B.
Moreover, we found that the effective temperature profile 
becomes significantly flatter in model A, 
but slightly steeper in model B, 
$T_{\rm eff} \propto r^{-p}$ with $p>1/2$.
Hence, the profile strongly depends on the 
distribution of viscous heating rates.

The deviations of our results from the slim disk
originate for the following reason.
In the slim disk,
using the vertically-integrated approximation,
the accreting velocity of the flow is investigated 
by solving the equation of motion in the $r$-direction,
and the radiative flux from the disk surface is determined
as a function of radius so as to satisfy 
the vertically integrated energy equation,
$Q_{\rm vis}=\sigma T_{\rm eff}^4+Q_{\rm adv}$,
where 
$Q_{\rm adv}$ is the advective cooling term.
The resultant luminosity increases 
in accordance with rise in the mass accretion rate,
and the effective temperature profile 
is obtained as $T_{\rm eff} \propto r^{-1/2}$
in the limit of large $\dot{m}$.
However, the radiative flux from the disk surface is assumed to 
be related to the temperature on the equatorial plane as
$\sigma T_{\rm eff}^4\sim \sigma T_{\rm c}^4/\tau$.
Since 
the radiation energy produced near the equatorial plane can not be 
radiated away immediately at the same radius 
but is advected inward at $r<r_{\rm trap}$,
the effective temperature in reality does not always reflect the temperature
of the equatorial plane at the same radius.
The relation of $\sigma T_{\rm eff}^4 \sim \sigma T_{\rm c}^4/\tau$
actually holds, only when the photon-trapping 
is not appreciable.
Hence, the slim disk cannot give the accurate radiative flux 
at the disk surface, 
although a part of the photon-trapping effects, the advection of the energy, 
is taken into consideration in the equation of energy balance.
%
In the present study, on the other hand,
we have actually solved radiative transfer in the $z$-direction
(in the comoving frame) 
to investigate the radiative flux from the disk surface.
Consequently, the observed luminosity and the effective temperature profile
deviate from the prediction of the slim disk.

%

\subsection{Future work}
We have so far 
assume that the disk size is practically infinite,
but the super-critical accretion disk would be much dimmer 
when the disk size, $r_{\rm out}$, is smaller than
the trapping radius.
In model A,
most of energy arises near equatorial plane of the disk,
and thus cannot arrive at the disk surface if $r_{\rm trap}>r_{\rm out}$.
As a result, the luminosity in such a case must be extremely small,
$L \ll L_{\rm E}$.
In model B,
the luminosity is estimated as 
\begin{eqnarray}
  L
  &\sim& 
  2 \int_{r_{\rm in}}^{r_{\rm out}} 
  2\pi r  Q_{\rm vis} \left( \frac{r}{r_{\rm trap}}\right)^{1/2}
  dr
  \nonumber\\
  &=& \frac{3}{4}\dot{M}_{\rm crit}c^2
  \left( \frac{2\dot{m}}{3h} \right)^{1/2}
  \nonumber\\
  & &
  \times \left[ 
  -2 \left( \frac{r_{\rm g}}{r_{\rm out}} \right)^{1/2}
  +\left( \frac{r_{\rm g}}{r_{\rm in}} \right)^{1/2}
  +\left( \frac{r_{\rm in}r_{\rm g}}{r_{\rm out}^2} \right)^{1/2}
  \right],
\end{eqnarray}
when $r_{\rm trap}>r_{\rm out} $.
For example,
this equation indicates
$L/L_{\rm E} \sim 3.1$ and $0.11$ for $r_{\rm out} = 10^2r_{\rm g}$
and $10r_{\rm g}$, respectively, 
in comparison with $L/L_{\rm E} \sim 30$ 
for the case of $r_{\rm out} \gg r_{\rm trap}$,
where we assume $r_{\rm in}=3r_{\rm g}$, $\dot{m}=10^4$, and $h=0.5$.
Such a situation might be realized 
in the accretion disks in the compact binaries,
in the central regions of Type II supernovae
(e.g., Mineshige et al. 1997)
and in gamma-ray bursts
(e.g., Narayan, Piran, \& Kumar 2001).

Throughout the present study,
we describe the simple model for accretion motion
without treating the hydrodynamical simulations.
If the radiative flux force at the disk surface 
is stronger than the gravity of the black hole,
the gas will be blown out like a wind
(Eggum, Coroniti, \& Katz 1987, 1988;
Okuda, Fujita, \& Sakashita 1997; Fujita \& Okuda 1998).
Since the ratio of the radiative flux force 
to the gravity in the $z$-direction is roughly estimated as 
$\sim \dot{m} (r/r_{\rm g})^{-1} F^z_{\rm surf}/Q_{\rm vis}$,
the photon-trapping effects tend to prevent outflow 
by suppression of $F^z_{\rm surf}/Q_{\rm vis}$.
Indeed, the gas would accrete onto the black hole
without being blown out in model A (see Figure 3).
Further study demands
multi-dimensional radiation-hydrodynamical simulation.

Moreover, we notice two very important improvement to be made.
One is to include convection effect,
which seems to occur when disk is radiation pressure-dominated
(Shakura, Sunyaev, \& Zilitinkevich 1978;
Agol et al. 2001).
Even if the viscous heating is effective in the vicinity of 
the equatorial plane as in model A,
the convection might more efficiently transfer the energy 
from the equatorial plane to the disk surface
than radiation.
Then, resultant properties of the disk would become close to model B,
but it is necessary for investigating in detail to solve
two dimensional hydrodynamical equations coupled 
with radiation transfer.
Second is to take the relativistic effect into consideration.
It would play a significant role around the inner edge of the disk.

\subsection{Observational implications}

There seem to be several 
possible sites of super-critical accretion flow,
where the photon-trapping effects may play a significant role.
Firstly, we mention micro-quasars which compose a peculiar subclass
of X-ray transients seemingly containing black holes
(see Tanaka \& Shibazaki 1996 for a review).
There is an indication that they emit rather high luminosity,
near Eddington.
In the case of GRS 1915+105, for example,
its black hole mass determination was done as $18.6\pm 2.2 M_\odot$ 
by Borozdin et al. (1999) based on the model fitting of the X-ray spectra.
Independently, Greiner, Cuby, \& McCaughrean (2001) obtained 
$14\pm 4 M_\odot$ 
through observations of Doppler shifted CO lines of the secondary star. 
Thus, its X-ray luminosity of several times $10^{39}$ erg s$^{-1}$ at the peak
is close to the Eddington luminosity of 
$(2.3\pm 0.3) \times 10^{39}$ erg s$^{-1}$.
Bursting behavior could be explained by relaxation oscillation
between the slim and standard disk
(Honma, Kato, \& Matsumoto 1991;
Yamaoka, Ueda, \& Inoue 2001).
If this is the case, this provides strong evidence for the presence
of a super-critical accretion flow.

Second one is ultra luminous X-ray sources (ULXs).
ULXs with intermediate-massive black holes are successively discovered in
nearby galaxies and some of them seem to emit large luminosity,
corresponding to the Eddington luminosity (Okada et al. 1998;
Colbert \& Mushotzky 1999; Mizuno et al. 1999; Makishima et al. 2000).
In addition, an intermediate massive black hole of $\gsim 10^3 M_\odot$
which is discovered in M82 (Matsumoto et al. 2001)
is also a good candidate of super-critical accretion,
although it is not always clear if their luminosities are really
close to the Eddington luminosities,
since there is no accurate mass estimations.

Narrow-line Seyfert 1 galaxies (NLS1s),
which are believed to have a central black hole,
are also considered to be sites of super-critical accretion
(Mineshige et al. 2000).
Since the NLS1s exhibit large soft X-ray excess
(Boller, Brandt, \& Fink 1996; 
Otani, Kii, \& Miya 1996; Leighly 1999; Hayashida 2000),
mass of the central black hole is estimated 
as $M \sim 10^{5-7}M_\odot$
by a comparison with soft X-ray bump in galactic black hole candidates.
Such moderate mass is also supported by reverberation mapping 
(Laor et al. 1997)
and X-ray variability method (Hayashida 2000),
and is consistent with narrow Balmer lines of NLS1s.
Thus, the observed luminosity,
$L \sim 10^{43-44}$, is near the Eddington luminosity,
and it implies that the accretion flow is super-critical.

The final candidate is gamma-ray bursts (GRBs)
(Piran 1999, M\'esz\'aros, Rees, \& Wijers 1999; M\'esz\'aros 2001).
Interestingly, many models (such as mergers of double neutron stars,
neutron star--black hole mergers, collapsars, and so on)
predict a similar configuration as a final product;
namely, a stellar-mass black hole surrounded by a
massive torus with a mass of 0.01 -- a few $M_\odot$.
Then, gas accretion onto a new-born black hole may be the origin of
huge energy release and high Lorentz factor
(e.g. Narayan, Paczy\'nski, \& Piran 1992; 
Narayan, Piran, \& Kumar 2001).
A simple estimation gives an enormous mass accretion rate.
If gas with mass of $M_{\rm gas}$ falls onto a black hole 
with a mass $M$ on a time-scale of $t_{\rm acc}$ sec, we find
${\dot M} \simeq 2 \times 10^{33} (M_{\rm gas}/M_\odot)
(t_{\rm acc}/1{\rm s})^{-1} {\rm g\, s^{-1}}$,
that is,
\begin{equation}
  {\dot m} = {\dot M}c^2/L_{\rm E}
  \sim 4\times 10^{15} 
  \left(\frac{M_{\rm gas}}{M_\odot} \right)
  \left(\frac{M}{3M_\odot} \right)^{-1}
  \left(\frac{t_{\rm acc}}{1 {\rm \, s}} \right)^{-1},
\end{equation}
(see, e.g., Narayan, Piran, \& Kumar 2001).
Then, the photon-trapping radius 
should be huge, $r_{\rm trap} \sim 10^{15}~r_{\rm g}$.
Since the disk size is, at most, about the size of giant stars, 
$r_{\rm out} \sim 10^{13-14} {\rm cm} \sim 10^{7-8}~
(M/3M_\odot)^{-1} r_{\rm g}$,
it is very likely that the emission from the entire disk surface is 
totally blocked as we already mentioned,
although neutrino energy loss may be essential in such a massive disk
(Ruffert \& Janka 1999; Popham, Woosley, \& Fryer 1999).

\section{CONCLUSIONS}
By employing a simple accretion flow model,
we have investigated the photon-trapping effects
in the super-critical black hole accretion flows
by solving radiation transfer and energy equations of 
gas as well as radiation.
The present results are summarized as follows.

(1) 
The radiative flux at the disk surface is reduced 
within $\sim \dot{m} r_{\rm g}$ due to photon trapping.
Then, the luminosity is kept around the Eddington luminosity
if the viscous heating is effective 
only in the vicinity of the equatorial plane.
Thus, the energy-conversion efficiency is drastically on decrease,
$\eta \equiv L/\dot{M}c^2 \sim \dot{m}^{-1}$.
Even if the energy is dissipated equally in the disk,
the energy-conversion efficiency is also suppressed 
in accordance with rise in the mass accretion rate,
$\eta \propto \dot{m}^{-1/2}$.

(2)
The slim disk can not accurately describe the 
effective temperature profile,
since it is assumed to be related to the temperature on the equatorial plane 
as $T_{\rm eff} \sim T_{\rm c}/\tau^{1/4}$.
As a result, the slim disk overestimates the luminosity of the disk
at $L\sim L_{\rm E}$.
The photon-trapping effects are already appreciable at the luminosities
as low as $\lsim L_{\rm E}$, while they only become critical at 
$L \gsim 3L_{\rm E}$ according to the slim disk calculations.

(3)
Through the photon-trapping effects,
the effective temperature profile becomes flatter
than $T_{\rm eff} \propto r^{-3/4}$
within the photon trapping radius in the super-critical accretion.
If the energy is dissipated in the vicinity of the equatorial plane,
the profile is almost flat 
($T_{\rm eff} \propto r^{-p}$ with $p \sim 0$)
and the maximum temperature decreases 
in accordance with the rise in the mass accretion rate.

(4)
The information regarding the 
distribution of the viscous heating rates, $q_{\rm vis}(z)$,
can be obtained through the observations of 
luminous X-ray objects at $L \gsim L_{\rm E}$.
In accordance with rise in the mass accretion rate,
the spectral index, $\zeta = d\log L_\nu / d\log \nu$ 
[0.1 keV -- 1.0 keV] for $M=10M_\odot$,
increases but in the different ways, depending the functional form
of $q_{\rm vis}(z)$.
When the gas is heated up uniformly, independently of the vertical
height, especially, 
the spectral index stays between $-0.5$ and 0,
although the index must be further small, $\ll -1$, 
in the case that the energy is dissipated around the equatorial plane.
Moreover, the most luminous radius shifts 
to the outer region as long as $\dot{m} \gg 10^2$ in the latter case.

\acknowledgments

We are grateful to J. Fukue, T. Nakamoto, and 
K. Watarai, for helpful discussion.
The calculations were carried out at 
Yukawa Institute for Theoretical Physics in
Kyoto University.
This work is 
supported in part by Research Fellowships of the Japan Society
for the Promotion of Science for Young Scientists, 02796 (KO)
and the Grants-in Aid of the
Ministry of Education, Science, Culture, and Sport, 
13640238 (SM), 09874055 (MU).

\end{document}